# TEMPERATURE DEPENDENCE OF THE TENSILE PROPERTIES OF SINGLE WALLED CARBON NANOTUBES: O(N) TIGHT BINDING MD SIMULATION

GÜLAY DERELİ [*], BANU SÜNGÜ

*Department of Physics, Yildiz Technical University, 34210 Istanbul, Turkey*

## Abstract

This paper examines the effect of temperature on the structural stability and mechanical properties of 20 layered (10,10) single walled carbon nanotubes (SWCNTs) under tensile loading using an O(N) tight-binding molecular dynamics (TBMD) simulation method. We observed that (10,10) tube can sustain its structural stability for the strain values of 0.23 in elongation and 0.06 in compression at 300K. Bond breaking strain value decreases with increasing temperature under stretching but not under compression. The elastic limit, Young's modulus, tensile strength and Poisson ratio are calculated as 0.10, 0.395 TPa, 83.23 GPa, 0.285, respectively, at 300K. In the temperature range from 300K to 900K; Young's modulus and the tensile strengths are decreasing with increasing temperature while the Poisson ratio is increasing. At higher temperatures, Young's modulus starts to increase while the Poisson ratio and tensile strength decrease. In the temperature range from 1200K to 1800K, the SWCNT is already deformed and softened. Applying strain on these deformed and softened SWCNTs do not follow the same pattern as in the temperature range of 300K to 900K.

*Corresponding author: gdereli@yildiz.edu.tr*

1. Introduction

Tensile properties of SWCNTs have been widely investigated by experimental and theoretical techniques. Experimentally, the Young's modulus of SWCNTs are measured as ranging from 0.9 to 1.9 TPa in [1]. In the SEM measurements of [2], SWCNT ropes broke at the strain values of 5.3 % or lower and the determined mean values of breaking strength and Young's modulus are 30 GPa and 1002 GPa, respectively. AFM and SGM measurements [3] show that SWCNTs can sustain elongations as great as 30% without breaking.

On the other hand, the *ab initio* simulation study of SWCNTs [4] showed that Young's modulus and Poisson ratio values of the tubes are ranging from 0.5 TPa to 1.1 TPa and from 0.11 to 0.19, respectively. The Young modulus and Poisson ratio of armchair nanotubes is given as 0.764 TPa and 0.32 respectively in [5]. Young modulus of (10,0); (8,4) and (10,10) tubes are calculated as 1.47 TPa; 1.10 TPa and 0.726 TPa, respectively in [6]. The results of [7] proposed that the structural failure should occur at 16% for zigzag and above 24% for armchair tubes. An empirical force-constant model of [8] gave the Young's modulus between 0.971 TPa- 0.975 TPa and Poisson ratio between 0.277 - 0.280. An empirical pair potential simulations of [9] gave the Young's modulus between 1.11 TPa -1.258 TPa and the Poisson ratio between 0.132-0.151. Continuum shell model of [10], calculated the elastic modulus as 0.94 TPa.and the maximum stress and failure strain values as 70 GPa, 88 GPa; 11%, 15% for (17,0) and (10,10) tubes, respectively. Finite element method [11] determined the strength of CNTs between 77 GPa to 101 GPa and Poisson ratio between 0.31-0.35. Analytical model in [12] found the tensile strength as 126.2 GPa of armchair tubes to be stronger than that (94.56 GPa) of zigzag tubes and the failure strains are 23.1% for armchair and 15.6-17.5% for zigzag tubes. MD simulations of [13-17] determined Young's modulus between 0.311 to 1.017 TPa for SWCNT. We found the Young's modulus, tensile strength and the Poisson ratio as 0.311 TPa, 4.92 GPA and 0.287 for (10,10) tubes in [16]. C.Goze et al. [18] calculated the Young's modulus as 0.423 TPa and Poisson ratio as 0.256 for (10,10) tube. Nonlinear elastic properties of SWCNTs under axial tension and compression were studied by T.Xiao et al. [19,20] using MD simulations with the second-generation Brenner potential. They showed that the energy change of the nanotubes are a cubic function of the tensile strains, both in tension and under compression. The maximum elongation strains are 15% and 17% for zigzag and armchair tubes, respectively. Also the maximum compression strain decreases with increasing tube diameter, and it is almost 4% for (10,10) tube. M.Sammalkorpi et al. [21] studied the effects of vacancy-related defects on the mechanical characteristics of SWCNTs by employing MD simulations and continuum theory. They calculated the Young's modulus for perfect

SWCNTs as 0.7 TPa. They showed that at 10K temperature, the critical strains of (5,5) and (10,10) tubes are 26% and 27%, respectively; also tensile strength is 120 GPa. On the other hand, for (9,9) and (17,0) tubes, the critical strains are found as 22% and 21%, respectively, and tensile strength is 110 GPa. Y.Wang et al. [22] investigated the compression deformation of SWCNTs by MD simulations using the Tersoff-Brenner potential to describe the interactions of carbon atoms. They determined that the SWCNTs whose diameters range from 0.5 nm to 1.7 nm and length ranges from 7 nm to 19 nm, the Young's modulus range from 1.25 TPa to 1.48 TPa. S.H.Yeak et al. [23] used MD and TBMD method to examine the mechanical properties of SWCNTs under axial tension and compression. Their results showed that the Young's modulus of the tubes are around 0.53 TPa; the maximum strain under axial tension is 20% for (12,12) and (7,7) tubes and also under this strain rate, the tensile stresses are 100 GPa and 90 GPa, respectively. Many elastic characteristics like the Young's modulus show a wide variations (0.3 TPa- 1.48 TPa) in all reported results in literature. These results are obtained at room temperature or without the mention of the temperature. The following reasons may be given for the variety of results: i) Young's modulus depends on the tube diameter and the chirality ii) different values are used for the wall thickness iii) different procedures are applied to represent the strain iv) accuracy of the applied methods (first principle methods in comparison with emprical model potentials)

SWCNTs will be locally subject to abrupt temperature increases in electronics circuits and the temperature increase affects their structural stability and the mechanical properties. MD simulation studies on the mechanical properties of the SWCNTs at various temperatures under tensile loading simulations can be followed in [24-**28**]. M.B.Nardelli et al. [24] showed that all tubes are brittle at high strains and low temperatures, while at low strains and high temperatures armchair nanotubes can be completely or partially ductile. In zigzag tubes ductile behavior is expected for tubes with n<14 while larger tubes are completely brittle. N.R.Raravikar et al. [25] showed between 0-800K temperature range radial Young's modulus of nanotubes decreases with increasing temperature and its slope is -7.5x$10^{-5}$ (1/K). C.Wei et al. [26,27] studied the tensile yielding of SWCNTs and MWCNTs under continuous stretching using MD simulations and a transition state theory based model. They showed that the yield strain decreases at higher temperatures and at slower strain rates. The tensile yield strain of SWCNT has linear dependence on the temperature and has a logarithmic dependence on the strain rate. The slope of the linear dependence increases with temperature. From their results it is shown that the yield strain of (10,0) tube decreased from 18% to 5% for the temperature range increasing from 300K to 2400K and for the different strain rates. Another

MD simulation study was performed by Y.-R.Jeng et al. [28] investigated the effect of temperature and vacancy defects on tensile deformation of (10,0); (8,3); (6,6) tubes of similar radii. Their Young's modulus and Poisson ratio values range from 0.92 to 1.03 TPa and 0.36-0.32, respectively. Their simulations also demonstrate that the values of the majority of the considered mechanical properties decrease with increasing temperature and increasing vacancy percentage.

In this study, the effect of temperature increase on the structural stability and mechanical properties of (10,10) armchair SWCNT under tensile loading is investigated by using O(N) tight-binding molecular dynamics (TBMD) simulations. Extensive literature survey is given in order to show the importance of our present study. The armchair 20 layers (10,10) SWCNT is chosen in the present work because it is one of the most synthesized nanotube in the experiments. For the first time we questioned how the strain energy of these nanotubes changes for positive and negative strain values at high temperatures. Along with the high temperature stress-strain curves for the first time we displayed the bond-breaking strain values through total energy graphs. Mechanical properties (Young's modulus, Poisson ratio, tensile strength and elastic limit) of this nanotube are reported at high temperatures.

## 2. Method

Traditional TB theory solves the Schrödinger equation by direct matrix diagonalization, which results in cubic scaling with respect to the number of atoms $O(N^3)$. The O(N) methods, on the other hand, make the approximation that only the local environment contributes to the bonding and hence the bond energy of each atom. In this case the run time would be in linear scaling with respect to the number of atoms. G.Dereli et al. [29,30] have improved and succesfully applied the O(N) TBMD technique to SWCNTs. In this work, using the same technique, we performed SWCNT simulations depending on conditions of temperature and unaxial strain. The electronic structure of the simulated system is calculated by a TB Hamiltonian so that the quantum mechanical many body nature of the interatomic forces is taken into account. Within a semi-empirical TB, the matrix elements of the Hamiltonian are evaluated by fitting a suitable database. TB hopping integrals, repulsive potential and scaling law is fixed in the program [31,32]. Application of the technique to SWCNTs can be seen in our previous studies [16,29,30].

An armchair (10,10) SWCNT consisting of 400 atoms with 20 layers is simulated. Periodic boundary condition is applied along the tube axis. Velocity Verlet algorithm along with the canonical ensemble molecular dynamics (NVT) is used. Our simulation procedure is as follows: i.) The tube is simulated at a specified temperature during a 3000 MD steps of run with a time step of 1 fs. This eliminates the possibility of the system to be trapped in a metastable state. We wait for the total energy per atom to reach the equilibrium state. ii.) Next, uniaxial strain is applied to the tubes. We further simulated the deformed tube structure (the under uniaxial strain) for another 2000 MD steps. In our study, while the nanotube is axially elongated or contracted, reduction or enlargement of the radial dimension is observed. Strain is obtained from $\varepsilon = (L - L_0)/L_0$, where $L_0$ and $L$ are the tube lengths before and after the strain, respectively. We applied the elongation and compression and calculated the average total energy per atom. Following this procedure we examined the structural stability, total energy per atom, stress-strain curves, elastic limit, Young's modulus, tensile strength, Poisson ratio of the (10,10) tube as a function of temperature.

The stress is determined from the resulting force acting on the tube per cross sectional area under stretching. The cross sectional area $S$ of the tube, is defined by $S = 2\pi R \delta R$, where $R$ and $\delta R$ are the radius and the wall thickness of the tube, respectively. We have used 3.4 Å for wall thickness. Mechanical properties are calculated from the stress-strain curves. Elastic limit is obtained from the linear regions of the stress-strain curves. Young's modulus, which shows the resistivity of a material to a change in its length, is determined from the slope of the stress-strain curve at studied temperatures. The tensile strength can be defined as the maximum stress which may be applied to the tube without perturbing its stability. Poisson ratio which is a measure of the radial reduction or expansion of a material under tensile loading can be defined as

$$\nu = -\frac{1}{\varepsilon}\left(\frac{R - R_o}{R_o}\right)$$

where $R$ and $R_o$ are the tube radius at the strain $\varepsilon$ and before the strain, respectively.

3. Results and Discussion

In Figure 1, we present the total energy per atom of the (10,10) SWCNT as a function of strain. Several strain values are applied. The positive values of strain corresponds to elongation and the negative values to compression. We obtained the total energy per atom vs strain curves in the temperature range between 300K-1800K in steps of 300K. Total energy per atom increases as we increase the temperature. An asymmetric pattern is observed in these curves. Repulsive forces are dominant in the case of compression. SWCNT does not have a high strength for compression as much as for elongation. (10,10) SWCNT is stable up to 0.06 strain in compression in the temperature range between 300K-1500K and 0.03 at 1800K. In elongation (10,10) SWCNT is stable up to 0.23 strain at 300K. As we increase the temperature the tube is stable up to 0.15 in elongation until 1800K. At 1800K we can only apply the strain of 0.08 in elongation before bond breakings. Figure 2a shows the variation of the total energy per atom during simulations for the strain values of 0.23 in elongation and 0.06 in compression at 300K. This figure indicates that the tube can sustain its structural stability up to these strain values. Beyond these, bond-breakings between the carbon atoms are observed at the strain values of 0.24 in elongation and 0.07 in compression as given in Figure 2b. In Figure 2b. sharp peaks represent the disintegrations of atoms from the tube. Next, bond-breaking strains are studied with increasing temperature. In Figure 3, we show the bond-breaking strain values with respect to temperature: as the temperature increases, disintegration of atoms from their places is possible at lower strain values due to the thermal motion of atoms. But this is not the case for compression as can be seen in Figure 3. Some examples of the variation of the total energy as a function of MD Steps under uniaxial strain values at various temperatures are given in Figure 4 and Figure 5. Figure 4a and Figure 5a shows that the tube can sustain its structural stability for strain values of 0.14 in elongation and 0.06 in compression at 900K; 0.08 in elongation and 0.03 in compression at 1800K, respectively. Beyond these points, bond-breakings between carbon atoms are observed at the strain values of 0.15 in elongation and 0.07 in compression, at 900 K (Figure 4b) and 0.09 in elongation and 0.04 in compression, at 1800K (Figure 5b).

The stress-strain curves of the tube are given in Figure 6. at studied temperatures. Our results show that the temperature have a significant influence on the stress-strain behaviour of the tubes. The stress-strain curves are in the order of increasing temperatures between 300K-900K. Stress value is increasing with increasing temperature . On the other hand between 1200K-1800K the stress value decreases with increasing temperature. This is due to the smaller energy difference under tensile loading with respect to 300K-900K temperature range. This result can also be followed in the total energy changes observed in Figures 2b , 4b and 5b.

Table1. gives a summary of the variations of the mechanical properties of (10,10) SWCNT with temperature. As given in Table1, elastic limit has the same value (0.10) in the 300K-900K temperature range. It drops to 0.09 in the 1200K-1500K temperature range and to 0.08 at 1800K. Young's modulus, Poisson ratio and the tensile strength of the tube have been found to be sensitive to the temperature (Table 1.). Our calculated value at 300K is 0.401 TPa.

It decreases to 0.370 TPa at 600K and to 0.352 TPa at 900K. In this temperature range Young's modulus decreases 12 %. After 1200K as we increase the temperature to 1800K there is 3% increase in the Young's modulus. We determined the tensile strength of (10,10) tube as 83.23 GPa at 300K. There is an abrupt decrease in tensile strength as we increase the temperature to 900K. Between 900K-1500K temperature range tensile strength does not change appreciably. At 1800K, it drops to 43.78 GPa. We specified the Poisson ratio at 300K as 0.3. Between 300K-900K temperature range Poisson ratio increases to 0.339 (12.5 %). This corresponds to the increase in the radial reduction. As we increase the temperature to 1200K its value drops to 0.315 and at 1800K to 0.289. We can conclude that for 20 layer (10,10) SWCNT in the 300K-900K temperature range : Young's modulus, the tensile strengths are decreasing with increasing temperature while the Poisson ratio is increasing. At higher temperatures, Young's modulus and the tensile strengths start to increase while the Poisson ratio decreases. In the 1200K-1800K temperature range, the SWCNT is already deformed and softened. Applying strain on these deformed and softened SWCNT do not follow the same pattern of 300K- 900K temperature range.

4. Conclusion

This paper reports for the first time the effect of temperature on the stress-strain curves, Young's modulus, tensile strength, Poisson ratio and elastic limit of (10,10) SWCNT. Total energy per atom of the (10,10) tube increases with axial strain under elongation. We propose that SWCNTs do not have a high strength for compression as much as for elongation. This is due to the dominant behavior of repulsive forces in compression. At room temperature, the bond breaking strain values of the tube are 0.24 in elongation and 0.07 in compression. We showed that as the temperature increases, the disintegration of atoms from their places is possible at lower strain values (0.09 at 1800K) in elongation due to the thermal motion of atoms. But this is not the case for compression. For 20 layers SWCNT bond-breaking negative strain values are temperature independent between 300K-1500K temperature range. Bond breakings occurs at 0.07 compression in this temperature range. When we increase the number of layers to 50, bond-breaking negative strain value decrease from 0.07 to 0.05 and remains the same in this temperature range. However, this is not a robust property for negative strains. When we decrease on the other hand the layer size to 10; bond-breaking negative strain values vary with increasing temperature. We note that for short tubes the critical strain values for compressive deformations are dependent on the size of the employed supercell and therefore they are an artifact of the calculation. In literature, various critical strain values were mentioned for the tube deformations. Our room temperature critical strain values are in aggrement with the experimental results of [3] and the computational results of [7,10,12]. MD simulations of [21] determined the critical strain value of (10,10) tube as 0.27 at 10K. To our knowledge the only reported temperature simulation study on tensile property comes from the MD simulation results of [26]. They showed that the yield strain of (10,0) tube decreases from 0.18 to 0.05 for the temperature range increasing from 300K to 2400K. Our results follow the same trend such that the bond breaking strain values decrease with increasing temperature. In [20] the maximum compression strain of (10,10) tube is given as 0.04 using Brenner potential without the mention of temperature, we obtained this value at 1800K.

We obtained the stress-strain curves in the temperature range between 300K-1800K. Our results show that the temperature have a significant influence on the stress-strain behavior of the tubes. (10,10) tube is brittle between 300K-900K and soft after 1200K. The elastic limit decreased from 0.10 to 0.08 with increasing temperature. There is a wide range of values given in literature for Young's modulus of SWCNTs due to the accuracy of the method and the choice of the wall thickness of the tube. The experimental results are in the range from 0.9 TPa to 1.9 TPa [1,2], *ab initio* results are in the range from 0.5 TPa to 1.47 TPa [4-6], empirical results are in the range from 0.971 TPa to 0.975 TPa [8] and from 1.11 TPa to 1.258 TPa [9], and also MD simulation results are in the range from 0.311 TPa to 1.48 TPa [13-28]. Our calculated value at 300K is 0.401 TPa is consistent with [4, 18,23]. We determined the tensile strength of (10,10) tube as 83.23 GPa at 300K and it decreases with increasing temperature. Maximum stress value of (10,10) tube is reported as 88 GPa in [10], 77 to 101 GPa in [11]. At 300K, we calculated the Poisson ratio of (10,10) tube as 0.3. This is in accord with the *ab initio* [5]; empirical [8,11], tight binding [18] results. M.B.Nardelli et al. [24] showed that all tubes are brittle at high strains and low temperatures, while at low strains and high temperatures armchair nanotubes can be completely or partially ductile. Our findings agree with this (10,10) armchair SWCNT is brittle at low temperatures and ductile at higher temperatures. Contrary to [28] our extensive temperature study has shown that Young's modulus changes with temperature.

5. Comments

Carbon nanotubes have the highest tensile strength of any material yet measured, with labs producing them at a tensile strength of 63 GPa, still well below their theoretical limit of 300 GPa. Carbon nanotubes are one of the strongest and stiffest materials known, in terms of their tensile stress and Young's modulus. This strength results from the covalent $sp^2$ bonds formed between the individual carbon atoms. Our simulation study using the interactions between electrons and ions also predicts a similar tensile strength and also shows that when exposed to heat they still keep their tensile strength around this value until very high temperatures like 1800K. CNTs are not nearly as strong under compression. Because of their hollow structure and high aspect ratio, they tend to undergo buckling when placed under compressive stress. The elastic limit is the maximum stress a material can undergo at which all strain are recoverable. (i.e., the material will return to its original size after removal of the stress). At stress levels below the elastic limit the material is said to be elastic. Once the material exceeds this limit, it is said to have undergone plastic deformation (also known as permanent deformation). When the stress is removed, some permanent strain will remain, and the material will be a different size. Our study shows that when the nanotube is exposed to heat this property does not change appreciably until 1800K. Through our tight-binding molecular dynamics simulation study we reported the high temperature positive/negative bond- breaking strain values and stress-strain curves of (10,10) SWCNTs . As far as we are aware, the strain energy values corresponding to positive/negative strain values at different temperatures are given here for the first time. We hope this extensive study of high temperature mechanical properties will be useful for aerospace applications of CNTs.

Acknowledgement

The research reported here is supported through the Yildiz Technical University Research Fund Project No: 24-01-01-04. The calculations are performed at the Carbon Nanotubes Simulation Laboratory at the Department of Physics, Yildiz Technical University, Istanbul, Turkey.

| Temperature (K) | Elastic Limit | Young's Modulus (TPa) | Tensile Strength (GPa) | Poisson Ratio |
| --- | --- | --- | --- | --- |
| 300 | 0.10 | 0.401 | 83.23 | 0.300 |
| 600 | 0.10 | 0.370 | 69.78 | 0.332 |
| 900 | 0.10 | 0.352 | 67.62 | 0.339 |
| 1200 | 0.09 | 0.360 | 67.33 | 0.315 |
| 1500 | 0.09 | 0.356 | 68.14 | 0.320 |
| 1800 | 0.08 | 0.365 | 43.78 | 0.289 |

Table 1. High Temperature Mechanical Properties of (10,10) SWCNT

Figure Captions

Figure 1. "Color online" Total energy per atom curves as a function of strain at different temperatures (negative strain values correspond to compression).

Figure 2 a. "Color online" (10,10) SWCNT is stable for the strains of 0.23 and -0.06 at 300K.

Figure 2 b. "Color online" Bond- breakings are observed between the carbon atoms for the strains of 0.24 and -0.07 at 300K. System is not in equilibrium.

Figure 3. "Color online" Bond- breaking strain variations as a function of temperature for a) tension, b) compression.

Figure 4 a. "Color online" (10,10) SWCNT is stable for the strains of 0.14 and -0.06 at 900K.

Figure 4 b. "Color online" Bond- breakings are observed between the carbon atoms for the strains of 0.15 and -0.07 at 900 K. System is not in equilibrium.

Figure 5 a. "Color online" (10,10) SWCNT is stable for the strains of 0.08 and -0.03 at 1800K.

Figure 5 b. "Color online" Bond- breakings are observed between the carbon atoms for the strains of 0.09 and -0.04 at 1800K. System is not in equilibrium.

Figure 6. "Color online" The stress-strain curves of (10,10) SWCNT at different temperatures.